\documentclass[intlimits,twoside,a4paper]{article}

\usepackage{amsmath,amssymb}
\usepackage{graphicx}

\usepackage[T2A]{fontenc}
\usepackage[cp1251]{inputenc}

\usepackage{amsthm}
\usepackage{emlines}

\usepackage[eqsecnum]{cmpj2}
\newtheorem{theorem}{Theorem}[section]
\newtheorem{proposition}{Proposition}[section]
\issue{2013}{16}{2}{23702}

\doinumber{10.5488/CMP.16.23702}

\title[A current algebra approach]%
{A current algebra approach to the equilibrium classical statistical
mechanics and its applications%
\thanks{Authors devote this work to their friend and colleague, a smart phase
transitions researcher Professor Mykhailo Kozlovskii on the occasion of his
60-years anniversary.}}
\author[N.N. Bogolubov, A.K. Prykarpatsky]{ N.N. Bogolubov (Jr.)\refaddr{label1}, A.K. Prykarpatsky\refaddr{label2,label3} }
\addresses
{\addr{label1} V.A.~Steklov Mathematical Institute of RAN, Moscow, Russian Federation
\addr{label2} AGH University of Science and Technology,  30-059 Krakow, Poland
\addr{label3}  Ivan Franko State Pedagogical University,  Drohobych, Ukraine}

\date{Received October 11, 2012}
\authorcopyright{N.N. Bogolubov, A.K. Prykarpatsky, 2013}

\begin{document}

\maketitle

\begin{abstract}  The non-relativistic current algebra approach is analyzed subject to its
application to studying the distribution functions of many-particle systems
at the temperature  equilibrium and their stability properties. We show
that  the classical  Bogolubov generating functional method is a very
effective tool for constructing the irreducible current algebra
representations and the corresponding different generalized measure
expansions including collective variables transform. The effective
Hamiltonian operator construction and its spectrum peculiarities subject to
the stability of  equilibrium many-particle systems are discussed.
\pacs 73.21.Fg, 73.63.Hs, 78.67.De
\end{abstract}

\section{Introduction}

It is well known \cite{Ar,Aref,Newm,CH} that a complete physical theory, both
relativistic and non-relativistic, can be described entirely in terms of
current algebra operators, such as current densities, rather than in terms
of canonical field operators, which is motivated by the fact that the current
densities are physically observable quantities contrary to the canonical
non-observable field operators. Moreover, the current algebra approach
appeared to be also very effective in studying both quantum and classical
statistical problems of many-particle systems by means of the Bogolubov
generating functional, whose mathematical structure became a fruitful source
of many approximation methods in modern statistical physics. Amongst them it
is necessary to mention a very powerful collective variables transform
suggested firstly by D.~Bohm \cite{Bom} and deeply developed by N.~Bogolubov \cite{Bo}, D.~Zubarev \cite{Zub} and I.~Yukhnovskii \cite{Yu}. This
transform has been reanalyzed in terms of the current algebra approach for
classical many-particle systems in \cite{BP,Pr,BP1}, where there was
constructed a corresponding Bogolubov generating functional of
distributions as a mathematical expectation of an infinite hierarchy of
the non-interacting many-particle systems embedded into an external
oscillatory potential field, with respect to a suitably defined infinite
divisible Gauss type measure.

Based on these results and making use of some additional properties of the
corresponding functional equations for the Bogolubov generating functional
of many-particle distribution functions we have constructed, for the case of
classical statistical mechanics, a new operator representation for an effective
Hamiltonian operator defined in a suitable Hilbert space, whose ground
state energy peculiarities make it possible to conceive the physical nature of the
related phase transitions and to describe the behavior of multi-particle
distribution functions.

%\part
\section
{Non-relativistic quantum and statistical mechanics: the current algebra
approach}

We assume a non-relativistic spinless many particle system of density $%
\bar{\rho}\in \mathbb{R}_{+}$ to be described by means of the non-relativistic
quantum Hamiltonian operator
\begin{equation}
\mathbf{H}:=\frac{\hbar^{2}}{2m}\underset{}{\int_{\mathbb{R}^{3}}}%
\rd^{3}x\langle\nabla \psi ^{+}(x),\nabla \psi (x)\rangle+\frac{1}{2}\underset{}{\int_{%
\mathbb{R}^{3}}\rd^{3}x\int_{\mathbb{R}^{3}}}\rd^{3}yW(x,y)\psi ^{+}(x)\psi
^{+}(y)\psi (y)\psi (x)  \label{2.1}
\end{equation}%
acting in a suitable Fock space $\Phi ,$ here we have denoted by $\langle\cdot
,\cdot \rangle$ the standard scalar product in the Euclidean space $\mathbb{R}%
^{3}, $ $W:\mathbb{R}^{3}\times \mathbb{R}^{3}\rightarrow \mathbb{R}$ is a
translation invariant interaction potential, and creation $\psi ^{+}(x):\Phi
\rightarrow \Phi $ $,$ $x\in \mathbb{R}^{3}$ and annihilation operators $%
\psi (y):\Phi \rightarrow \Phi$, $y\in \mathbb{R}^{3},$ satisfy the
standard canonical commutation relationships:
\begin{equation}
\lbrack \psi (y),\psi ^{+}(x)]=\delta (x-y), \qquad [\psi ^{+}(x),\psi
^{+}(y)]=0=[\psi (x),\psi (y)].  \label{2.2}
\end{equation}

The current algebra representation of the Hamiltonian operator (\ref{2.1})
\ is based on the following self-adjoint density operators: particle number
density%
\begin{equation}
\rho (x):=\psi ^{+}(x)\psi (x)  \label{2.3}
\end{equation}%
and current density%
\begin{equation*}
J(x):=\frac{1}{2\ri}[\psi ^{+}(x)\nabla \psi (x)-\nabla \psi ^{+}(x)\psi (x)]
\end{equation*}%
at any point $x\in \mathbb{R}^{3},$ satisfying the well known classical
current Lie algebra commutator relationships:%
\begin{eqnarray}
\lbrack \rho (f_{1}),\rho (f_{2})] \, =0, \qquad [\rho (t),J(g)]=\ri\rho
(\langle g,\nabla f\rangle ),  \label{2.4} \qquad
\lbrack J(g_{1}),J(g_{2})] =\ri J([g_{2},g_{1}]),
\end{eqnarray}%
where we have defined the smeared \cite{CH,MS,Vl} density operators%
\begin{equation}
\rho (f):=\int_{\mathbb{R}^{3}}\rd^{3}x\rho (x)f(x), \qquad J(g):=\int_{%
\mathbb{R}^{3}}\rd^{3}x\langle g(x),J(x)\rangle  \label{2.5}
\end{equation}%
for any Schwartz functions $f\in \mathcal{S}(\mathbb{R}^{3};\mathbb{R})$ and
$g\in \mathcal{S}(\mathbb{R}^{3};\mathbb{R}^{3})$ and $%
[g_{2},g_{1}]:=\langle g_{2},\nabla \rangle g_{1}$ $-\langle g_{1},\nabla \rangle g_{2}$ for any $%
g_{1},g_{2}\in J(\mathbb{R}^{3};\mathbb{R}^{3}).$

The following proposition characterizes \cite{Ar,CH,MS,MS1} the current Lie
algebra  (\ref{2.4}) from the group representation theory.

\label{Prop_2.1} The exponential current operators are as follows:%
\begin{equation}
U(f):=\exp [\ri\rho (f)],\qquad V(\varphi _{t}^{g}):=\exp [\ri tJ(g)],
\label{2.5a}
\end{equation}%
where $\rd\varphi _{t}^{g}(x)/\rd t:=g\circ \varphi _{t}^{g}(x)$ and $g\circ
\varphi _{t}^{g}(x):=g[\varphi _{t}^{g}(x)]$ for any $t\in \mathbb{R}$ and $%
x\in \mathbb{R}^{3},$ satisfy the current group relationships%
\begin{equation}
U(f_{1})U(f_{2})=U(f_{1}+f_{2}),\qquad V(\varphi )U(f)=U(f\circ \varphi
)V(\varphi ),\qquad V(\varphi _{1})\circ V(\varphi _{2})=V(\varphi
_{2}\circ \varphi _{1})  \label{2.6}
\end{equation}%
for the semi-simple product $G:=\mathcal{S}\rtimes \mathrm{Diff}(\mathbb{R}^{3})$ and
the abelian Schwartz group $\ \mathcal{S}(\mathbb{R}^{3};\mathbb{R}),$where $%
f_{1},f_{2}$ and $f\in \mathcal{S}(\mathbb{R}^{3};\mathbb{R}),$ $\varphi
_{1},\varphi _{2}$ and $\varphi \in \mathrm{Diff}(\mathbb{R}^{3}).$ The latter
appeared to be very important in constructing the corresponding group $G=%
\mathcal{S}\rtimes \mathrm{Diff}(\mathbb{R}^{3})$ representations in suitable Hilbert
spaces and their physical interpretation as a classical generating Bogolubov
functional \cite{Bo,MS,MS1,BP1} for the corresponding many-particle
distribution functions.

The Hamiltonian operator  (\ref{2.1}) permits the following current algebra
representation
\begin{equation}
\mathbf{H=}\frac{\hbar ^{2}}{8m}\int_{\mathbb{R}^{3}}\rd^{3}x\langle K^{+}(x),\rho
^{-1}(x)K(x)\rangle+\frac{1}{2}\int_{\mathbb{R}^{3}}\rd^{3}x\int_{\mathbb{R}%
^{3}}\rd^{3}yW(x,y):\rho (x)\rho (y): \,,  \label{2.7}
\end{equation}%
where, by definition,
\begin{equation}
K(x):=\nabla \rho (x)+2\ri J(x)  \label{2.8}
\end{equation}%
for any $x\in \mathbb{R}^{3}$ and the normal ordering \cite{BB,BS} acts as
\begin{equation}
:\rho (x_{1})\rho (x_{2})\ldots\rho (x_{n}):= \prod_{j=1}^{n}\left[\rho
(x_{j})-\sum_{k=1}^{j-1}\delta (x_{j}-x_{k})\right]  \label{2.9}
\end{equation}%
for arbitrary $x_{j}\in \mathbb{R}^{3},j=\overline{1,n},n\in \mathbb{Z}_{+}.$

The current group $G=\mathcal{S}\rtimes \mathrm{Diff}(\mathbb{R}^{3}),$ as is well
known, possesses many different irreducible unitary representations in
suitable Hilbert spaces. In particular, in the standard $N$-particle Hilbert
space $\mathcal{H}^{(N)}:=L_{2}^{(\mathrm{sym})}(\mathbb{R}^{3N};\mathbb{C})$  for
an arbitrary but fixed $N\in \mathbb{Z}_{+}$  the particle density operator
acts as
\begin{equation}
\rho (f)\omega =\Big[\sum_{j=1}^{N}f(x_{j})\Big]\omega   \label{2.9a}
\end{equation}%
and the current density operator acts as
\begin{equation}
J(g)\omega =\frac{1}{2\ri}\sum_{j=1}^{N}\left[\langle g(x_{j}),\nabla _{j}\rangle + \langle \nabla
_{j},g(x_{j})\rangle \right]\omega  \label{2.10}
\end{equation}%
for any $f\in \mathcal{S}(\mathbb{R}^{3};\mathbb{R}),g\in \mathcal{S}(%
\mathbb{R}^{3};\mathbb{R}^{3})$ and arbitrary vector $\omega \in \hat{H}%
^{(N)}.$

In the general case, the current group  (\ref{2.6}) possesses many different
irreducible unitary representations in suitable Hilbert spaces $\mathcal{H},$
which can be written down as
\begin{equation}
\mathcal{H=}\int_{\mathcal{S}^{\prime }}^{\oplus }\rd\mu (F)\mathcal{H}_{F}\,,
\label{2.11}
\end{equation}%
where $\mu :2^{\mathcal{S}^{\prime }}\rightarrow \mathbb{R}_{+}$ is some
cylindrical measure on the generalized space $\mathcal{S}^{\prime }:=%
\mathcal{S}^{\prime }(\mathbb{R}^{3};\mathbb{R}),$ $\mathcal{H}_{F}$ are
marked by elements $F\in \mathcal{S}^{\prime }(\mathbb{R}^{3};\mathbb{R})$
complex linear spaces, which for many physical applications \cite{Newm,MS,CH}
are one-dimensional. In the case $\dim \mathcal{H}_{F}=1$, one obtains from
(\ref{2.11}) that $\mathcal{H}\ \simeq L_{2}^{(\mu )}(\mathcal{S}^{\prime
};\mathbb{C}).$ Now, if an element $\omega (F)\in \mathcal{H}$ is taken
arbitrarily, from  (\ref{2.6}) one  easily follows that
\begin{eqnarray}
U(f)\omega (F) &=&\exp [\ri(F,f)]\omega (F)\,,  \label{2.12} \notag\\
V(\varphi )\omega (F) &=&\chi _{\varphi }(F)\omega (\varphi ^{\ast }F)\left[
\frac{\rd\mu (\varphi ^{\ast }F)}{\rd\mu (F)}\right] ^{1/2},
\end{eqnarray}%
where, by definition, $(\varphi ^{\ast }F,f):=(F,f\circ \varphi ),$ ${%
\rd\mu (\varphi ^{\ast }F)}/{\rd\mu (F)}$ is the standard Radon-Nykodym
derivative of the measure $\mu (\varphi ^{\ast }F)$ with respect to the
measure $\mu (\varphi ^{\ast }F)$ and $\chi _{\varphi }(F)$ is a
complex-valued factor of a unit norm, referred to as the co-cycle, satisfying the
relationship%
\begin{equation}
\chi _{\varphi _{2}}(F)\chi _{\varphi _{1}}(\varphi _{2}^{\ast }F)=\chi
_{\varphi _{1}\circ \varphi _{2}}(F)  \label{2.13}
\end{equation}%
for any $\varphi _{1},\varphi _{2}\in \mathrm{Diff}(\mathbb{R}^{3})$ and arbitrary
point $F\in \mathcal{S}^{\prime }(\mathbb{R}^{3};\mathbb{R}).$

Now, based on the expression  (\ref{2.12}), one can define the
following functional%
\begin{equation}
\mathcal{L}(f):=(\Omega ,\exp [\ri\rho (f)]\Omega )  \label{2.14}
\end{equation}%
for the \textit{physical} ground state vector $\Omega \in \mathcal{H}$ of
the suitably renormalized \cite{Ar,MS,MS1,BP1} Hamiltonian operator  (\ref%
{2.7}):%
\begin{equation}
\Omega :=\arg \inf_{\omega \in \mathcal{H},||\omega ||=1}\frac{(\omega ,%
\mathbf{H}\omega )}{(\omega ,\mathbf{N}\omega \mathbf{)}}\,.  \label{2.15}
\end{equation}%
This, in particular, means that the Hilbert space representation of the
current group $G=\mathcal{S}\rtimes \mathrm{Diff}(\mathbb{R}^{3})$ is chosen in such
a way that the corresponding Hamiltonian operator  (\ref{2.7}) is bounded
from below, thereby realizing a respectively stable physical many-particle
system. Taking this into account one can put, without loss of generality,
that the Hamiltonian operator  (\ref{2.7}) and its ground state vector  (%
\ref{2.15}) satisfy the equivalent conditions
\begin{equation}
\mathbf{H}\Omega =0, \qquad(\Omega ,\rho (x)\Omega )=\bar{\rho}%
>0\,.  \label{2.16}
\end{equation}

Now, we can interpret the functional  (\ref{2.14}) as a generating
functional of the current group $G=\mathcal{S}\rtimes \mathrm{Diff}(\mathbb{R}^{3})$
irreducible representations \cite{MS,MS1,BP} in the physically proper
Hilbert space $\mathcal{H}.$ This is based on the following theorem \cite{Ar}
owing to H.~Araki.

\begin{theorem}
\label{Tm_2.2}
A functional $\mathcal{L}:G\rightarrow \mathbb{C}$ generates a unitary representation of the group $G$ if and only if there
exists a unitary continuous representation $\pi :G\rightarrow Aut$ $\mathcal{%
H}$ with a cyclic vector $\Omega \in \mathcal{H}$ satisfying the condition%
\begin{equation}
\mathcal{L}(a)=(\Omega ,\pi (a)\Omega ),\qquad \mathcal{H}=\mathrm{span}\{a\Omega
\in \mathcal{H}:a\in G\}  \label{2.17}
\end{equation}%
for any $a\in G.$
\end{theorem}

Having applied theorem \ref{Tm_2.2} to  the functional (\ref{2.14}), we
derive that by means of constructing suitable generating functionals subject
to the given Hamiltonian operator  (\ref{2.7}) one can find the
corresponding operator representations of the current group $G=\mathcal{S}%
\rtimes \mathrm{Diff}(\mathbb{R}^{3})$ and vice versa.

\section{The current algebra representations and the Hamiltonian operator
reconstruction}

Based on the relationships  (\ref{2.12}) and the generating functional
expression  (\ref{2.14}), one can easily calculate that
\begin{equation}
\mathcal{L}(f)=\int_{\mathcal{S}^{\prime }}\exp \{\ri(F,f)\}\rd\mu (F)
\label{3.1}
\end{equation}%
for some suitably determined quasi-invariant measure $\mu :2^{\mathcal{S}%
^{\prime }}\rightarrow \mathbb{R}_{+},$ that is an ergodic measure with
respect to the diffeomorphism group $\mathrm{Diff}(\mathbb{R}^{3})$: for any $\mathrm{Diff}(%
\mathbb{R}^{3})$-invariant set $Q\subset \mathcal{S}^{\prime }(\mathbb{R}%
^{3};\mathbb{R})$ either $\mu (Q)=0$ or $\mu (\mathcal{S}^{\prime
}\backslash Q)=0.$ As a result of  (\ref{3.1}) one finds, as an example,
that the standard quantum mechanical $N$-particle representation of the
current group $G=\mathcal{S}\rtimes \mathrm{Diff}(\mathbb{R}^{3})$ is described \cite%
{Ar,MS,BP} by the generalized singular measure
\begin{equation}
\rd \mu (F)=\Omega ^{\ast }\Omega \prod\limits_{j=1}^{N}\rd^{3}x_{j}\delta
\left(F-\sum_{k=1}^{N}\delta (x-x_{k})\right),  \label{3.2}
\end{equation}%
whose support  \textrm{supp} $\mu =\{F\in \mathcal{S}^{\prime }(\mathbb{R}^{3};%
\mathbb{R}):F=\sum_{k=1}^{N}\delta (x-x_{k})\}.$

Consider now the generating functional  (\ref{2.14}) and observe that the
following quantities
\begin{eqnarray}
F_{n}(x_{1},x_{2},\ldots,x_{n}) &:=&(\Omega ,:\rho (x_{1})\rho (x_{2})\ldots\rho
(x_{n}):\Omega ) \label{3.3} \\
&=&:\frac{1}{\ri}\frac{\delta }{\delta f(x_{1})}\frac{1}{\ri}\frac{\delta }{%
\delta f(x_{2})}\ldots\frac{1}{\ri}\frac{\delta }{\delta f(x_{n})}:\mathcal{L}%
(f)\Big|_{f=0}  \notag
\end{eqnarray}%
for arbitrary $n\in \mathbb{Z}_{+}$ represent the $n$-particle distribution
functions of the quantum mechanical many-particle system with the
Hamiltonian  (\ref{2.7}), or equivalently, the functional  (\ref{2.14})
is respectively, the Bogolubov generating functional of many-particle
distribution functions. Since the essence of the Bogolubov generating
functional is held in the correspondingly derived \cite{Bo} functional equation, we proceed now to determine its exact analytical form
taking into account the structure of the related current group $G=\mathcal{S}%
\rtimes \mathrm{Diff}(\mathbb{R}^{3})$ representation in a suitable Hilbert space $%
\mathcal{H}.$

Following the standard operator construction, suggested in \cite{MS,MS1},
one can define  a selfadjoint operator $A(x;\rho ):\mathcal{H\rightarrow H}%
,\ x\in \mathbb{R}^{3},$ by means of the relationships%
\begin{equation}
K(g)\Omega =A(g;\rho )\Omega ,  \label{3.4}
\end{equation}%
satisfied for any $g\in \mathcal{S}(\mathbb{R}^{3};\mathbb{R}^{3}),$ where
we put, by definition,
\begin{equation}
K(g):=\int_{\mathbb{R}^{3}}\rd^{3}x\langle g(x),K(x)\rangle ,\qquad A(g;\rho ):=\int_{%
\mathbb{R}^{3}}\rd^{3}x\langle g(x),A(x;\rho )\rangle.  \label{3.5}
\end{equation}%
To proceed further we need an important proposition concerning the matrix
elements of the operators $J(g)$ and $\mathbf{H}:\mathcal{H\rightarrow H}$
for any $g\in \mathcal{S}(\mathbb{R}^{3};\mathbb{R}^{3}).$

\begin{proposition}
\label{Prop_3.1}
Let $\{\tilde{f}=\exp[\ri\rho (f)]\Omega \in \mathcal{H}%
:f\in \mathcal{S}(\mathbb{R}^{3};\mathbb{R})\mathcal{\}}$  be the set of
vectors dense in the Hilbert space $\mathcal{H}$ owing to the Araki's
theorem \ref{Tm_2.2}. Then, the following scalar product expressions%
\begin{eqnarray}
(\tilde{f}_{1},J(g)\tilde{f}_{2}) &=&(\tilde{f}_{1},\rho (\langle g,\nabla (\tilde{f%
}_{1}+\tilde{f}_{2})\rangle)\tilde{f}_{2}),  \label{3.6}\notag \\
(\tilde{f}_{1},\mathbf{H}\tilde{f}_{2}) &=&\frac{\hbar ^{2}}{8m}(\tilde{f}%
_{1},\rho (\langle \nabla \tilde{f}_{1},\nabla \tilde{f}_{2}\rangle)\tilde{f}_{2})
\end{eqnarray}%
hold for all $f_{1},f_{2}\in \mathcal{S}(\mathbb{R}^{3};\mathbb{R}).$
\end{proposition}

Based now on simple enough but slightly cumbersome calculations, one can
derive the following renormalized Hamiltonian operator expression:%
\begin{equation}
\mathbf{\tilde{H}=}\frac{\hbar ^{2}}{8m}\int_{\mathbb{R}^{3}}\rd^{3}x\langle \tilde{K}%
^{+}(x;\rho ),\rho ^{-1}(x)\tilde{K}(x;\rho )\rangle,  \label{3.7}
\end{equation}%
where, by definition, the operator
\begin{equation}
\tilde{K}(x;\rho ):=K(x)-A(x;\rho )  \label{3.7a}
\end{equation}%
satisfies the condition
\begin{equation}
\tilde{K}(x;\rho )\Omega =0  \label{3.8}
\end{equation}%
for all $x\in \mathbb{R}^{3}$. Now, making use of  (\ref{3.6}), we can
rewrite the defining condition  (\ref{3.8}) in the following functional
equation form:%
\begin{equation}
\lbrack \nabla _{x}-\nabla f(x)]\frac{1}{\ri}\frac{\delta \mathcal{L}(f)}{%
\delta f(x)}=A(x;\delta )\mathcal{L}(f),  \label{3.9}
\end{equation}%
where we put for any $x\in \mathbb{R}^{3}$%
\begin{equation}
A(x;\delta ):=A(x;\rho )\big|_{\rho =\frac{1}{\ri}\frac{\delta }{\delta f}}\,.
\label{3.10}
\end{equation}%
Similarly, one can calculate the matrix element values for the renormalized
Hamiltonian operator  (\ref{3.7}) subject to the irreducible cyclic
representation of the current group $G=\mathcal{S}\rtimes \mathrm{Diff}(\mathbb{R}%
^{3})$:%
\begin{equation}
(\tilde{f}_{1},\mathbf{\tilde{H}}\tilde{f}_{2})=\frac{\hbar ^{2}}{8m}(\tilde{%
f}_{1},\rho (\langle\nabla \tilde{f}_{1},\nabla \tilde{f}_{2}\rangle)\tilde{f}_{2})=(%
\tilde{f}_{1},\mathbf{H}\tilde{f}_{2})  \label{3.11}
\end{equation}%
for all $f_{1},f_{2}\in \mathcal{S}(\mathbb{R}^{3};\mathbb{R}),$ meaning
that two current algebra operator representations  (\ref{2.7}) and  (\ref{3.7}) of the initial Hamiltonian operator  (\ref{2.1}), defined in the
canonical Fock space, are physically completely equivalent.

\section{The generating Bogolubov functional equation for
the temperature equilibrium states}

Assume that a classical many-particle system is at a bounded inverse
temperature $\beta \in \mathbb{R}_{+}$ and its Gibbs statistical operator
\begin{equation}
\mathcal{P}:=\frac{\exp (-\beta \mathbf{H)}}{\mathrm{tr}\exp(-\beta \mathbf{H})}\,,
\label{4.1}
\end{equation}%
where ``$\mathrm{tr}$'' means the standard trace-operation well determined on the
ideal of nuclear operators in a Hilbert space $\mathcal{H},$ in which the
corresponding irreducible unitary representation of the current group $G=%
\mathcal{S}\rtimes \mathrm{Diff}(\mathbb{R}^{3})$ is realized. To determine it
analytically, we define the Bogolubov generating functional of many-particle
distribution functions as
\begin{equation}
\mathcal{L}(f):=\mathrm{tr}\{\mathcal{P}\exp [\ri\rho (f)]\}  \label{4.2}
\end{equation}%
for any $f\in \mathcal{S}(\mathbb{R}^{3};\mathbb{R}).$ having imposed on
the functional  (\ref{4.2}) the Araki's conditions of theorem \ref{Tm_2.2},
we can easily derive that there exists \cite{Ar,Newm,CH,MS1,BP1} an
effective normalized cyclic vector $\Omega _{\beta }\in \mathcal{H}_{\beta
}, $ naturally corresponding to the effective Hamilton operator
\begin{equation}
\mathbf{\tilde{H}}_{\beta }:=\frac{\hbar ^{2}}{8m}\int_{\mathbb{R}%
^{3}}\rd^{3}x\langle\tilde{K}_{\beta }^{+}(x;\rho ),\rho ^{-1}(x)\tilde{K}_{\beta
}(x;\rho )\rangle ,  \label{4.3}
\end{equation}%
such that
\begin{equation}
\mathcal{L}(f)=(\Omega _{\beta },\exp [\ri\rho (f)]\Omega _{\beta }),
\label{4.4}
\end{equation}%
where we have put, by definition,
\begin{equation}
\tilde{K}_{\beta }(x;\rho ):=K(x)-A_{\beta }(x;\rho ),\qquad K(x)\Omega
_{\beta }:=A_{\beta }(x;\rho )\Omega _{\beta }  \label{4.5}
\end{equation}%
for any $x\in \mathbb{R}^{3}.$ As a result of the definition  (\ref{4.2})
and relationships  (\ref{4.5}), one  easily finds \cite{BP}, as the Planck
constant $\hbar \rightarrow 0,$ that
\begin{equation}
\mathcal{L}(f)=\frac{\exp [-\beta W(\delta )]\mathcal{L}_{0}(f)}{\exp
[-\beta W(\delta )]\mathcal{L}_{0}(f)|_{f=0}}\,,  \label{4.6}
\end{equation}%
where $\mathcal{L}_{0}(f),f\in \mathcal{S}(\mathbb{R}^{3};\mathbb{R}),$ is
the generating functional for the noninteracting equilibrium many-particle
system and, by definition, we put
\begin{equation}
W(\delta ):=W(\rho )\big|_{\rho =\frac{1}{\ri}\frac{\delta }{\delta f}}\,.
\label{4.7}
\end{equation}

Similarly to the  above reasonings one also finds that the generating
functional  (\ref{4.6}) satisfies \cite{Bo,BP} the Bogolubov type
functional equation
\begin{equation}
\lbrack \nabla _{x}-\nabla f(x)]\frac{1}{\ri}\frac{\delta \mathcal{L}(f)}{%
\delta f(x)}=A_{\beta }(x;\delta )\mathcal{L}(f)\,,  \label{4.8}
\end{equation}%
where the corresponding operator $A_{\beta }(x;\rho ):\mathcal{H}_{\beta
}\rightarrow \mathcal{H}_{\beta }$ for any $x\in \mathbb{R}^{3}$
linearly depends on the binary interparticle interaction potential $W:\mathbb{R}%
^{3}\times \mathbb{R}^{3}\rightarrow \mathbb{R}.$

Hence, one easily infers that the generating functional $\mathcal{L}%
_{0}(f)$, $f\in \mathcal{S}(\mathbb{R}^{3};\mathbb{R})$ for the unitary
representation of the current group $G=\mathcal{S}\rtimes \mathrm{Diff}(\mathbb{R}%
^{3})$ satisfies the reduced functional equation
\begin{equation}
\lbrack \nabla _{x}-\nabla f(x)]\frac{1}{\ri}\frac{\delta \mathcal{L}_{0}(f)}{%
\delta f(x)}=0\,,  \label{4.9}
\end{equation}%
whose general non-normalized solution equals the integral%
\begin{equation}
\mathcal{L}_{0}(f)=\int_{\mathbb{R}}\rd\mu _{\beta }(z)\exp \left(z\int_{\mathbb{R}%
^{3}}\rd^{3}x\left\{\exp[\ri f(x)]-1\right\}\right)  \label{4.10}
\end{equation}%
with respect to some Radon measure $\mu _{\beta }:2^{\mathbb{R}}\rightarrow
\mathbb{R}$ on the real axis $\mathbb{R}.$ Thus, submitting  (\ref{4.10})
into  (\ref{4.6}), we obtain from  (\ref{4.8}) that  for any $x\in
\mathbb{R}$
\begin{equation}
A_{\beta }(x;\rho )=-\beta \int_{\mathbb{R}^{3}}\rd^{3}y\nabla _{x}W(x,y):\rho
(x)\rho (y):  \label{4.11}
\end{equation}%
and, respectively,
\begin{equation}
\lbrack \nabla _{x}-\nabla f(x)]\frac{1}{\ri}\frac{\delta \mathcal{L}(f)}{%
\delta f(x)}=-\beta \int_{\mathbb{R}^{3}}\rd^{3}y\nabla _{x}W(x,y):\frac{1}{%
\ri}\frac{\delta }{\delta f(x)}\frac{1}{\ri}\frac{\delta }{\delta f(y)}:\mathcal{%
L}(f)\,.  \label{4.12}
\end{equation}

The functional equation (\ref{4.12}), being well known long ago owing to
the classical results of Bogolubov \cite{Bo}, makes it possible, using the current
algebra approach, to interpret it as an equation for the generating
functional of irreducible current group $G=\mathcal{S}\rtimes \mathrm{Diff}(\mathbb{R}%
^{3})$ representations in a suitable Hilbert space $\mathcal{H}_{\beta }$
with a cyclic vector $\Omega _{\beta }\in \mathcal{H}_{\beta }$ being the
ground state vector for a respectively renormalized ``\textit{effective}''
positive definite Hamiltonian operator  (\ref{4.3}) and satisfying the
conditions  (\ref{4.5}). This gives rise to the following canonical current
algebra representation of the Hamiltonian operator  (\ref{4.3}):%
\begin{eqnarray}
%\begin{array}{c}
\mathbf{\tilde{H}}_{\beta }&=&\frac{\hbar ^{2}}{8m}\int_{\mathbb{R}%
\!\!^{3}}\rd^{3}x\langle K^{+}(x),\rho ^{-1}(x)K(x)\rangle \nonumber\\
&&+\sum_{n-2\in \mathbb{Z}_{+}}\frac{1}{n!}\int_{\mathbb{R}^{3}}\!\!\rd^{3}x_{1}%
\int_{\mathbb{R}^{3}}\!\!\rd^{3}x_{2}\ldots\int_{\mathbb{R}^{3}}\!\!\rd^{3}x\tilde{W}%
_{\beta }^{(n)}(x_{1},x_{2},\ldots,x_{n}):\rho (x_{1})\rho (x_{2})\ldots\rho
(x_{n}):\,,%
%\end{array}
\label{4.13}
\end{eqnarray}%
where the effective inter-particle potentials $\tilde{W}_{\beta }^{(n)}:%
\underset{j=1}{\overset{n}{\times }}\mathbb{R}_{j}^{3}\rightarrow \mathbb{R}%
, $ $n-2\in \mathbb{Z}_{+},$  non-locally depend on the initial interparticle
potential  $W:\mathbb{R}^{3}\times \mathbb{R}^{3}\rightarrow \mathbb{R}$
and on the inverse temperature parameter $\beta \in \mathbb{R}_{\div }.$

The Hamiltonian operator  (\ref{4.13}) can be respectively transformed to
the canonical form%
\begin{eqnarray}
%\begin{array}{c}
\mathbf{\tilde{H}}_{\beta }&=&\frac{\hbar ^{2}}{2m}\int_{\mathbb{R}%
^{3}}\rd^{3}x\langle \nabla \psi ^{+}(x),\nabla \psi (x)\rangle\nonumber \\
&&+\sum_{n-2\in \mathbb{Z}_{+}}\frac{1}{n!}\int_{\mathbb{R}^{3}}\rd^{3}x_{1}%
\int_{\mathbb{R}^{3}}\rd^{3}x_{2}\ldots\int_{\mathbb{R}^{3}}\rd^{3}x\tilde{W}%
_{\beta }^{(n)}(x_{1},x_{2},\ldots,x_{n}) \nonumber\\
&&\times :\psi ^{+}(x_{1})\psi (x_{1})\psi ^{+}(x_{2})\psi (x_{2})\ldots\psi
^{+}(x_{n})\psi (x_{n}):%
%\end{array}
\label{4.14}
\end{eqnarray}%
acting in the standard Fock space $\Phi .$ The latter can be used for
determining the related canonical $N$-particle representation of the
Hamiltonian  (\ref{4.14}) by means of the following defining relationship:%
\begin{eqnarray}
%\begin{array}{c}
&&\mathbf{\tilde{H}}_{\beta }\left( \sum_{n\in \mathbb{Z}_{+}}\frac{1}{n!}%
\int_{\mathbb{R}^{3}}\rd^{3}x_{1}\int_{\mathbb{R}^{3}}\rd^{3}x_{2}\ldots\int_{%
\mathbb{R}^{3}}d^{3}xf_{n}(x_{1},x_{2},\ldots,x_{n}):|x_{1},x_{2},\ldots x_{n}\rangle%
\right) = \nonumber\\
&&= \sum_{n\in \mathbb{Z}_{+}}\frac{1}{n!}\int_{\mathbb{R}^{3}}\rd^{3}x_{1}%
\int_{\mathbb{R}^{3}}\rd^{3}x_{2}\ldots\int_{\mathbb{R}^{3}}\rd^{3}x\mathcal{H}%
_{\beta }^{(n)}f_{n}(x_{1},x_{2},\ldots,x_{n}):|x_{1},x_{2},\ldots x_{n}\rangle,%
%\end{array}
\label{4.15}
\end{eqnarray}%
where $f_{n}\in L_{2}^{(\mathrm{sym})}(\mathbb{R}^{3n};\mathbb{C}),n\in \mathbb{Z}%
_{+},$ and we put, by definition,
\begin{equation}
|x_{1},x_{2},\ldots x_{n}\rangle:=\prod\limits_{j=1}^{n}\psi ^{+}(x_{j})|0\rangle
\label{4.16}
\end{equation}%
the independent orthogonal states in the Fock space $\Phi$, generated by a
cyclic vacuum vector $|0\rangle\in \Phi $, satisfying the annihilation condition $%
\psi (x)|0\rangle=0$ for all $x\in \mathbb{R}^{3}$.

Having constructed the corresponding $N$-particle translational-invariant
Hamiltonian operators $\mathcal{H}_{\beta }^{(N)}:L_{2}^{(\mathrm{sym})}(\mathbb{R}%
^{3N};\mathbb{C})$ $\rightarrow L_{2}^{(\mathrm{sym})}(\mathbb{R}^{3N};\mathbb{C})$
for arbitrary $N\in \mathbb{Z}_{+}$ particles in a volume $\Lambda \subset
\mathbb{R}^{3},$ one can  study its \textit{a priori} positive spectrum $%
\sigma (\mathcal{H}_{\beta }^{(N)})\subset \mathbb{R}_{+}$ and its
peculiarities as a function of the density
\begin{equation}
\bar{\rho}=\mathrm{tr}\left\{\mathcal{P}\rho (x)\right\}=\frac{1}{\ri}\left. \frac{\delta \mathcal{L}%
(f)}{\delta f(x)}\right\vert _{f=0}  \label{4.17}
\end{equation}%
and the inverse temperature parameter $\beta \in \mathbb{R}_{+}.$ In
particular, the condition  (\ref{4.17}) allows one to determine the
above introduced measure $\rd\mu _{\beta },$ entering the functional  (\ref{4.10}).

Now, it is important to recall that an equilibrium many-particle statistical
system is stable \cite{Leb,Leb1}, if its generating functional satisfies the
well Kubo-Martin-Schwinger analycity condition. This condition, in
particular, imposes a strong analytical dependence on the inverse
temperature parameter $\beta \in \mathbb{R}_{+}$ of the spectrum $\sigma (%
\mathcal{H}_{\beta }^{(N)})\subset \mathbb{R}_{+}$  as $N\rightarrow \infty
$ in such a way that the density $\bar{\rho}=\lim N/\Lambda \in \mathbb{R}%
_{+}$ persists to be constant. One obtains another inference  from the
important fact that the number operator $\mathbf{N}_{\beta }\mathbf{:=}\int_{%
\mathbb{R}^{3}}\rd^{3}x\rho (x)$ is a conserved quantity, that is
\begin{equation}
\lbrack \mathbf{\tilde{H}}_{\beta },\mathbf{N}_{\beta }\mathbf{]=0}
\label{4.18}
\end{equation}%
for those parameters $\beta \in \mathbb{R}_{+},$ for which the equilibrium
many-particle system is stable and does not pass a phase transition.

\section{The current algebra representation aspects of the collective
variables transform}

The collective variables transform \cite{Bom,Bo,Yu,Pr,BP} allows one to
consequently take into account and separate two different impacts of a binary
interaction potential $W:=W^{(\mathrm{l})}+W^{(\mathrm{s})}$into the many-particle
distribution functions subject to  its long distance $W^{(\mathrm{l})}$ and short
distance $W^{(s)\text{ }}$ parts. Since the long distance interaction
potential  responds for the so-called ``collective''  behavior of the
many-particle at a bounded inverse temperature parameter $\beta \in \mathbb{R%
}_{+},$ the corresponding Bogolubov generating functional  (\ref{4.2}) can
be formally rewritten in the operational form as
\begin{eqnarray}
\mathcal{L}(f) &=&Z(f)/Z(0)\,,  \label{4.19} \notag\\
Z(f) &:=&\exp [-\beta W^{(\mathrm{s})}(\delta )]\mathcal{L}^{(\mathrm{l})}(f)\,,
\end{eqnarray}%
where, by definition,
\begin{equation}
\mathcal{L}^{(\mathrm{l})}(f):=\exp[-\beta W^{(\mathrm{l})}(\delta )]\mathcal{L}_{0}(f)\,.
\label{4.20}
\end{equation}%
The functional  (\ref{4.20}) can be quite easily calculated
by means of the Fourier transform representation of the long distance
interaction potential and the standard quasi-classical limit $\hbar
\rightarrow 0$:
\begin{eqnarray}
%\begin{array}{c}
\mathcal{L}^{(\mathrm{l})}(f)&=&\lim_{\hbar \rightarrow 0}\mathrm{tr}\Bigg\{ \mathcal{P}_{0}\exp
\Bigg[-\frac{\beta }{2}\int_{\mathbb{R}^{3}}\rd^{3}k\nu ^{(\mathrm{l})}(k):\rho _{k}\rho
_{-k}:\Bigg]\exp[\ri\rho (f)]\Bigg\} \nonumber\\
&=&\lim_{\hbar \rightarrow 0}\mathrm{tr}\Bigg\{ \mathcal{P}_{0}\exp \Bigg[-\frac{\beta }{2}%
\int_{\mathbb{R}^{3}}\rd^{3}k\int_{\mathbb{R}^{3}}\rd^{3}x\rho (x)\Bigg]\nonumber
 \\
&&\times \int_{\mathbb{C}}D(\omega )\exp \Bigg[-\int_{\mathbb{R}^{3}}\rd^{3}k\frac{%
2\pi ^{2}}{\beta \nu ^{(\mathrm{l})}(k)}\omega _{k}\omega _{-k}- \int_{\mathbb{R%
}^{3}}\rd^{3}k2\pi \ri\omega _{k}\rho _{k}\Bigg]\exp [\ri\rho (f)]\Bigg\} \nonumber \\
&=&\int_{\mathbb{C}}D(\omega )J(\omega )\lim_{\hbar \rightarrow 0}\mathrm{tr}\Bigg(
\mathcal{P}_{0}\exp \Bigg\{ \ri\rho \Bigg[ f-2\pi \int_{\mathbb{R}%
^{3}}\rd^{3}k\omega _{k}\exp\{\ri\langle k,x \rangle\}-\frac{\ri\beta }{2}\int_{\mathbb{R}%
^{3}}\rd^{3}k\nu ^{(\mathrm{l})}(k)\Bigg] \Bigg\} \Bigg) \nonumber \\
&=&\int_{\mathbb{R}}\rd\mu _{\beta }^{(\mathrm{l})}(\bar{z})\int_{\mathbb{C}^{\infty
}}D(\omega )J^{(\mathrm{l})}(\omega ;\bar{z})\exp \Bigg(\bar{z}\int_{\mathbb{R}%
^{3}}\rd^{3}x\Bigg\{\exp [\ri f(x)]-1\Bigg\}g(x;\omega )\Bigg) ,%
%\end{array}
\label{4.21}
\end{eqnarray}%
where we denoted the measure $D(\omega ):=\prod_{k\in \mathbb{R}^{3}}%
\frac{\ri}{2}\rd\omega _{k}\wedge \rd\omega _{-k}$, the parameter $\bar{z}:=z\exp
\big[-\frac{\beta }{2}\int_{\mathbb{R}^{3}}\rd^{3}k\nu ^{(\mathrm{l})}(k)\big]$, and the
free particle system statistical operator is equal to
\begin{equation}
\mathcal{P}:=\frac{\exp (-\beta \mathbf{H}_{0}\mathbf{)}}{\mathrm{tr}
\exp(-\beta
\mathbf{H}_{0})}\,, \qquad \mathbf{H}_{{0}}:=\frac{\hbar ^{2}%
}{2m}\int_{\mathbb{R}^{3}}\rd^{3}x\langle \nabla \psi ^{+}(x),\nabla \psi (x)\rangle \,,
\label{4.21a}
\end{equation}%
the Fourier transform
\begin{equation}
\nu ^{(\mathrm{l})}(k)=\frac{1}{(2\pi )^{3}}\int_{\mathbb{R}^{3}}\rd^{3}xW^{(\mathrm{l})}(x,y)%
\exp (\ri\langle k,x-y\rangle)  \label{4.21b}
\end{equation}%
and the measure kernels (``Jacobian'')
\begin{eqnarray}
J(\omega ) &:=& \exp \Bigg\{ -\int_{\mathbb{R}^{3}}\rd^{3}k\frac{2\pi ^{2}}{%
\beta \nu ^{(\mathrm{l})}(k)}\omega _{k}\omega _{-k}+\int_{\mathbb{R}^{3}}\rd^{3}k\ln
\frac{\pi }{\beta \nu ^{(\mathrm{l})}(k)}\Bigg\} ,  \notag \\
J^{(\mathrm{l})}(\omega ;\bar{z}) &:=& J(\omega )\exp \Bigg\{ \bar{z}\int_{\mathbb{R}%
^{3}}\rd^{3}xg^{(\mathrm{l})}(x;\omega )\Bigg\} ,  \label{4.22} \notag\\
g^{(\mathrm{l})}(x;\omega ) &:=& \exp \Bigg\{ -2\pi \ri\int_{\mathbb{R}^{3}}\rd^{3}k\omega
_{k}\exp (\ri\langle k,x \rangle )\Bigg\} .
\end{eqnarray}%
The formal series expansion%
\begin{eqnarray}
%\begin{array}{c}
J^{(\mathrm{l})}(\omega ;\bar{z})&=&J(\omega )\exp \Bigg\{ -\frac{\bar{z}(2\pi
)^{2}(2\pi )^{3}}{2}\int_{\mathbb{R}^{3}}\rd^{3}k\omega _{k}\omega
_{-k} \nonumber \\
&&+\sum_{n\in \mathbb{Z}_{+}\backslash \{2\}}\frac{\bar{z}(-2\pi \ri)^{n}(2\pi
)^{3}}{n!}\int_{\mathbb{R}^{3}}\rd^{3}k_{1}\int_{\mathbb{R}^{3}}\rd^{3}k_{2}\ldots%
\int_{\mathbb{R}^{3}}\rd^{3}k_{n}\prod\limits_{j=1}^{n} \omega
_{k_{j}}\delta \Bigg(\sum_{j=1}^{n}k_{j}\Bigg)\Bigg\} \nonumber \\
&=&\exp \Bigg\{ -\int_{\mathbb{R}^{3}}\rd^{3}k\frac{(2\pi )^{2}[\beta \nu
^{(\mathrm{l})}(k)(2\pi )^{3}\bar{z}+1]}{\beta \nu ^{(\mathrm{l})}(k)}\omega _{k}\omega
_{-k}+\ldots\Bigg\},%
%\end{array}
\label{4.23}
\end{eqnarray}%
owing to  (\ref{4.21})  and (\ref{4.23}), right away gives rise to the
approximation of the generating functional  (\ref{4.20}) by means of the so
called ``screened''  long distance potential
\begin{eqnarray}
\bar{W}^{(\mathrm{l})}(x,y) &:=&\int_{\mathbb{R}^{3}}\rd^{3}k\frac{\nu ^{(\mathrm{l})}(k)}{1+\nu
^{(\mathrm{l})}(k)\beta \bar{z}(2\pi )^{3}}\exp (\ri\langle k,x-y\rangle ) \notag \\
&=&\int_{\mathbb{R}^{3}}\rd^{3}k\bar{\nu}^{(\mathrm{l})}(k)\exp (\ri\langle k,x-y\rangle )
\label{4.24}
\end{eqnarray}%
under the external effect of an infinite set of oscillatory potentials:%
\begin{eqnarray}
\mathcal{L}^{(\mathrm{l})}(f) &=&\int_{\mathbb{R}}\rd\mu _{\beta }^{(\mathrm{l})}(\bar{z})\int_{%
\mathbb{C}^{\infty }}D(\omega )\exp \Bigg\{ -\int_{\mathbb{R}^{3}}\rd^{3}k%
\frac{(2\pi )^{2}}{\beta \bar{\nu}^{(\mathrm{l})}(k)}\omega _{k}\omega
_{-k}+\ldots\Bigg\}   \label{4.25}  \notag\\
&&\times \exp \Bigg( \bar{z}\int_{\mathbb{R}^{3}}d^{3}x\{\exp
[\ri f(x)]-1\}g^{(\mathrm{l})}(x;\omega )\Bigg).
\end{eqnarray}%
Having substituted the functional expression  (\ref{4.25}) into  (\ref%
{4.19}) one can easily obtain the corresponding Bogolubov generating
functional in the Ursell-Mayer type infinite expansion form, based on the
following operator expression:%
\begin{eqnarray}
Z(f) &:=&\exp [-\beta W^{(\mathrm{s})}(\delta )]\mathcal{L}^{(\mathrm{l})}(f)  \label{4.26}
\notag\\
&=&\exp [-\beta W^{(\mathrm{s})}(\delta )]\int_{\mathbb{R}}\rd\mu _{\beta }^{(\mathrm{l})}(%
\bar{z})\exp \Bigg( \sum_{n\in \mathbb{Z}_{+}}\frac{\bar{z}^{n}}{n!}\int_{%
\mathbb{R}^{3}}\rd^{3}x_{1}\int_{\mathbb{R}^{3}}\rd^{3}x_{2}\ldots\int_{\mathbb{R}%
^{3}}\rd^{3}x_{n}   \notag \\
&&\times  \prod\limits_{j=1}^{n}\{\exp
[\ri f(x_{j})]-1\}g_{n}^{(\mathrm{l})}(x_{1},x_{2},\ldots,x_{n})\Bigg) ,
\end{eqnarray}%
where $g_{n}^{(\mathrm{l})}:\mathbb{R}^{3n}\rightarrow \mathbb{R},n\in \mathbb{Z}%
_{+}$ are so-called $n$-particle ``cluster'' distribution functions.

Observe also that the Bogolubov type generating functional  (\ref{4.21})
can be rewritten in the integral Gauss type form as
\begin{eqnarray}
\mathcal{L}^{(\mathrm{l})}(f) &=&\int_{\mathbb{C}^{\infty }\times \mathbb{R}}\rd\mu
_{\beta }^{(\mathrm{l})}(\bar{z},\omega )J^{(\mathrm{l})}(\omega ,\bar{z})\exp \Bigg( \bar{z}%
\int_{\mathbb{R}^{3}}\rd^{3}x\{\exp [\ri f(x)]-1\}g^{(\mathrm{l})}(x;\omega )\Bigg)
\notag \\
&=&\int_{\xi \in \mathbb{C}^{\infty }\times \mathbb{R}}\rd\bar{\mu}_{\beta
}^{(\mathrm{l})}(\xi )\mathcal{L}_{0}^{(\mathrm{l})}(f;\xi ),  \label{4.27}
\end{eqnarray}%
where, by definition, $\xi :=(\omega ,\bar{z},\mathbb{R}^{3})$ and for any
Lebesgue measurable set $A\subset \mathbb{R\times C}^{\infty }\times \mathbb{%
R}^{3}$, the  measure
\begin{equation}
\bar{\mu}_{\beta }^{(\mathrm{l})}(A):=\int_{A\subset \mathbb{R\times C}^{\infty
}\times \mathbb{R}^{3}}\rd^{3}k\rd\mu _{\beta }^{(\mathrm{l})}(\bar{z})D(\omega
)J^{(\mathrm{l})}(\omega ,\bar{z})  \label{4.28}
\end{equation}%
with a reduced generating functional $\mathcal{L}_{0}^{(\mathrm{l})}(f;\omega ,\bar{%
z},k),(\omega ,\bar{z},k)\in \mathbb{C}^{\infty }\times \mathbb{R\times R}%
^{3},$ satisfying \cite{MS} the following functional equation:%
\begin{equation}
\lbrack \nabla _{x}-\nabla f(x)]\frac{1}{\ri}\frac{\delta }{\delta f(x)}%
\mathcal{L}_{0}^{(\mathrm{l})}(f;\omega ,\bar{z},k)=2\pi k\omega _{k}\exp (\ri\langle k,x \rangle )%
\frac{1}{\ri}\frac{\delta }{\delta f(x)}\mathcal{L}_{0}^{(\mathrm{l})}(f;\omega ,\bar{z}%
,k).  \label{4.29}
\end{equation}%
Hence, one can obtain the effective partial renormalized long distance
Hamiltonian operator%
\begin{equation}
\mathbf{\tilde{H}}_{\beta }^{(\mathrm{l})}(\omega ,\bar{z},k):=\frac{\hbar ^{2}}{8m}%
\int_{\mathbb{R}^{3}}\rd^{3}x\langle \tilde{K}_{\beta }^{(l)+}(x,\omega ,\bar{z}%
,k),\rho ^{-1}(x)\tilde{K}_{\beta }^{(\mathrm{l})}(x,\omega ,\bar{z},k)\rangle,
\label{4.30}
\end{equation}%
where we put, by definition,%
\begin{equation}
\tilde{K}_{\beta }^{(l)+}(x,\omega ,\bar{z},k):=K(x)-2\pi k\omega _{k}\rho
(x)\exp (\ri\langle k,x\rangle )  \label{4.31}
\end{equation}%
for all $x\in \mathbb{R}^{3}$ and $(\omega ,\bar{z},k)\in \mathbb{C}^{\infty
}\times \mathbb{R\times R}^{3}.$ As a result of  (\ref{4.30}), one finds
\cite{MS} that the effective external long distance oscillatory potential
\begin{equation}
\tilde{W}_{\beta }^{(\mathrm{l})}(x;\omega ,\bar{z},k)=\frac{\hbar ^{2}\pi \omega
_{k}k^{2}}{2m}\exp (\ri\langle k,x\rangle)[1+\pi \omega _{k}k\exp (\ri\langle k,x\rangle )]  \label{4.32}
\end{equation}%
yields the following partial canonical Hamiltonian operator expression:%
\begin{equation}
\mathbf{\tilde{H}}_{\beta }(\omega ,\bar{z},k)=\frac{\hbar ^{2}}{2m}\int_{%
\mathbb{R}^{3}}\rd^{3}x\langle \nabla \psi ^{+}(x),\nabla \psi (x)\rangle+\int_{\mathbb{R}%
^{3}}\rd^{3}x\tilde{W}_{\beta }^{(\mathrm{l})}(x;\omega ,\bar{z},k)\psi ^{+}(x)\psi (x),
\label{4.33}
\end{equation}%
whose generating functional of the current group $G=\mathcal{S}\rtimes \mathrm{Diff}(%
\mathbb{R}^{3})$ equals the expression  (\ref{4.27}). Moreover, taking into
account the representation  (\ref{4.26}), one can construct the full
effective long distance Hamiltonian operator%
\begin{equation}
\mathbf{\tilde{H}}_{\beta }^{(\mathrm{l})}=\frac{\hbar ^{2}}{8m}\int_{\mathbb{R}%
^{3}}\rd^{3}x\langle \tilde{K}_{\beta }^{(l)+}(x;\rho ),\rho ^{-1}(x)\tilde{K}_{\beta
}^{(\mathrm{l})}(x;\rho )\rangle,  \label{4.34}
\end{equation}
where, by definition, we put%
\begin{eqnarray}
%\begin{array}{c}
\tilde{K}_{\beta }^{(l)+}(x;\rho )&:=&K(x)-\sum_{n\in \mathbb{Z}_{+}}\frac{1}{%
(n-1)!}\int_{\mathbb{R}^{3}}\rd^{3}x_{1}\int_{\mathbb{R}^{3}}\rd^{3}x_{2}\ldots%
\int_{\mathbb{R}^{3}}\rd^{3}x_{n}\nonumber\\
&&\times \nabla _{x}\ln g_{n}^{(\mathrm{l})}(x_{1},x_{2},\ldots x_{n}):\rho (x_{1})\rho
(x_{2})\ldots\rho (x_{n}):\,.%
%\end{array}
\label{4.35}
\end{eqnarray}%
As a simple corollary from the expression  (\ref{4.34}), one obtains that
the effective long-distance Hamiltonian operator contains an infinite
hierarchy of  multinary potential energy terms, which should be in due
course taken into account when studying the peculiarities of the
corresponding many-particle distribution functions. In particular, the energy
spectrum of the $N$-particle canonical representation of the Hamiltonian
operator (\ref{4.34}) possesses an important information on the many-particle
system stability at a special inverse temperature parameter $\beta \in
\mathbb{R}_{+}.$

As an example demonstrating the effective Hamiltonian operator construction,
we will consider a one-dimensional many-particle system of density $\bar{\rho}%
\in $ $\mathbb{R}_{+}$ on an axis $\mathbb{R}$ at a finite inverse
temperature $\beta \in \mathbb{R}_{+},$ described by the following operator
expression in the canonical Fock space $\Phi$:
\begin{equation}
\mathbf{H=}\frac{\hbar ^{2}}{2m}\int_{\mathbb{R}}\rd x\langle \nabla \psi
^{+}(x),\nabla \psi (x)\rangle -\lambda \int_{\mathbb{R}}\rd x\int_{\mathbb{R}}\rd y\ln
|x-y|\psi ^{+}(x)\psi ^{+}(y)\psi (y)\psi (x),  \label{4.36}
\end{equation}%
where $\lambda \in \mathbb{R}_{+}$ is a positive parameter. The
corresponding Bogolubov generating functional $\mathcal{L}(f),$ $f\in
\mathcal{S}(\mathbb{R};\mathbb{R}),$ satisfies, owing to  (\ref{4.12}), the
following functional equation:%
\begin{equation}
\lbrack \nabla _{x}-\ri\nabla f(x)]\frac{1}{\ri}\frac{\delta \mathcal{L}(f)}{%
\delta f(x)}=2\lambda \beta \int_{\mathbb{R}}\rd y\nabla _{x}\ln \left\vert
x-y\right\vert :\frac{1}{\ri}\frac{\delta }{\delta f(x)}\frac{1}{\ri}\frac{%
\delta }{\delta f(x)}:\mathcal{L}(f)\,.  \label{4.37}
\end{equation}%
Taking now into account that the expression
\begin{equation}
A(x;\rho )=2\lambda \beta \int_{\mathbb{R}}\rd y\nabla _{x}\ln \left\vert
x-y\right\vert :\rho (x)\rho (y):  \label{4.38}
\end{equation}%
one can easily construct the effective renormalized Hamiltonian operator
\begin{equation}
\mathbf{\tilde{H}}_{\beta }\mathbf{:=}\frac{\hbar ^{2}}{8m}\int_{\mathbb{R}%
}\rd x\langle \tilde{K}_{\beta }^{+}(x;\rho ),\rho ^{-1}(x)\tilde{K}_{\beta }(x;\rho
)\rangle ,  \label{4.39}
\end{equation}%
acting in a suitable Hilbert space $\mathcal{H}_{\beta },$ realizing a
non-reducible representation of the basic current group $G=\mathcal{S}\rtimes
\mathrm{Diff}(\mathbb{R}).$ It is quite easy to calculate the resulting effective
Hamiltonian operator  (\ref{4.39}) expression in the canonical Fock space $%
\Phi$:%
\begin{equation}
\mathbf{\tilde{H}}_{\beta }\mathbf{=}\frac{\hbar ^{2}}{2m}\int_{\mathbb{R}%
}\rd x\langle\nabla \psi ^{+}(x),\nabla \psi (x)\rangle+\frac{\hbar ^{2}}{2m}\int_{\mathbb{R%
}}\rd x\int_{\mathbb{R}}\rd y\frac{\lambda \beta (\lambda \beta -1)}{|x-y|^{2}}%
\psi ^{+}(x)\psi ^{+}(y)\psi (y),\psi (x),  \label{4.40}
\end{equation}%
describing an infinite set of particles on the axis $\mathbb{R},$ binarily
interacting to each other by means of the inverse square potential%
\begin{equation}
\tilde{W}_{\beta }(x,y):=\frac{\hbar ^{2}\lambda \beta (\lambda \beta -1)}{%
2m|x-y|^{2}}\,,  \label{4.41}
\end{equation}%
where $x\neq y\in \mathbb{R}$ and $\lambda ,\beta \in \mathbb{R}_{+}$ are
suitable positive parameters. Here, it is necessary to mention that the
effective Hamiltonian operator  (\ref{4.40})  realizes a nonreducible
representation of the current group $G=\mathcal{S}\rtimes \mathrm{Diff}(\mathbb{R})$
in the Hilbert space $\mathcal{H}_{\beta },$ generated by its ground cyclic
eigenstate $\Omega _{\beta }\in \mathcal{H}_{\beta },$ satisfying the
determining condition  (\ref{2.15}).  It can be shown \cite{MS,MS1,BP1}
that%
\begin{equation}
\bar{\varepsilon}_{\beta }:=\underset{\omega \in \mathcal{H}_{\beta
},\left\Vert \omega \right\Vert =1}{\inf }\frac{(\omega ,\mathbf{\tilde{H}}%
_{\beta }\omega )}{(\omega ,\mathbf{\tilde{N}}_{\beta }\omega )}=\lambda
^{2}\beta ^{2}\pi ^{2}\bar{\rho}^{2}/6  \label{4.42}
\end{equation}%
holds, where we denoted by $\mathbf{\tilde{N}}_{\beta }:=\int_{\mathbb{R}%
}\rd x\rho (x)$ the corresponding particle number operator in the Hilbert
space $\mathcal{H}_{\beta }.$ The least average energy per particle  (\ref%
{4.42}) analytically depends on the inverse temperature parameter $\beta \in \mathbb{R}%
_{+}$. The same can also be obtained for the other energy
excitations of the Hamiltonian operator (\ref{4.40}). Thus, we infer that
the initial one-dimensional many-particle system with the Hamiltonian  (\ref%
{4.36}) and at the inverse temperature parameter $\beta \in \mathbb{R}_{+}$
is completely stable and permits no phase transition. Moreover, at the
temperature parameter $\beta =1/\lambda \in \mathbb{R}_{+}$ the effective
Hamiltonian operator (\ref{4.40}) describes a many-particle noninteracting
system of the density $\bar{\rho}\in \mathbb{R}_{+}$ and the least average
energy per particle $\bar{\varepsilon}_{\beta }=\pi ^{2}\bar{\rho}^{2}/6,$
depending only on the density.

\section{Conclusion}

The investigation of statistical properties of classical many-particle
systems at a finite inverse temperature $\beta \in \mathbb{R}_{+}$ and a
fixed density $\bar{\rho}\in \mathbb{R}_{+}$ by means of the current algebra
representations has two main reasons: firstly, it provides an interesting
reformulation of the initial quantum statistical problem in terms of
physical observables such as the particle number density and the particle
flux density, rather than the corresponding second-quantized field creation
and annihilation operators.

The second reason is related to a very rich structure of the current group
$G=\mathcal{S}\rtimes \mathrm{Diff}(\mathbb{R}^{3})$ irreducible representations,
according to the Bogolubov functional equation for the generating
many-particle distribution functional, and whose analytical property subject
to the temperature parameter $\beta \in $ $\mathbb{R}_{+}$ are responsible
for the system stability as it follows from the Kubo-Martin-Schwinger approach,
applied to the classical statistical mechanics.

Moreover, a very rich functional-operator structure of solutions to the
related Bogolubov functional equations allows one to make physically reasonable
re-expansions of the general irreducible representation measure, as it was
show for the case of the classical collective variables transform, and whose
generating functional permits an additive Gauss type representation, based on
an infinite set of free noninteracting many-particle systems embedded in an
external oscillatory type potential field.

As a dual aspect of irreducible representations of the current group $G=%
\mathcal{S}\rtimes \mathrm{Diff}(\mathbb{R}^{3}),$ related to the Bogolubov
functional equation, we need to mention the construction of associated
effective Hamiltonian operators subject to the basic ground state cyclic
representation of the current group, whose analytical properties are
responsible for the many-particle system stability and possibly, for the
phase transition behavior. We hope that the approach devised in the work will
prove to be helpful in further gaining insight into the statistical
clustering properties of many-particle systems and in developing new more
powerful and specialized analytical techniques for solving other interesting
problems in statistical physics.

\section{Acknowledgements}

Authors are cordially thankful to prof. J.~S{\l}awianowski and
prof. J.~Spa{\l}ek for interesting discussions, important comments and remarks.

\ukrainianpart

\title{Підхід до класичної рівноважної статистичної механіки на основі  алгебри струмів  та його застосування}
\author{М.М. Боголюбов (мол.)\refaddr{label1}, А.К. Прикарпатський\refaddr{label2,label3}}
\addresses
{\addr{label1} Математичний інститут ім. В.А. Стєклова РАН, Москва, Росія
\addr{label2} Академія гірництва та металургії, Краків, Польща
\addr{label3} Державний педагогічний університет ім. І. Франка,  Дрогобич, Україна
}

%
%% якщо автор є один або автори є з однієї установи:
%
%  \author{1й Автор, 2й Автор, \ldots}
%  \address{Інститут\ldots}
%
%%

\makeukrtitle

\begin{abstract}
\tolerance=3000%
Аналізується підхід до вивчення функцій розподілу систем багатьох частинок при рівноважній температурі та  властивості їх стабільності, що ґрунтується на представленнях нерелятивістичної   алгебри струмів. Показано, що метод породжуючого функції розподілу     класичного  функціоналу Боголюбова є досить ефективним інструментом для побудови незвідних представлень   алгебри  струмів  та      відповідних узагальнених    розкладів мір, включаючи відоме перетворення до  колективних змінних. Запропонована конструкція ефективного  оператора Гамільтона, обговорюються   особливості його спектра
в залежності від стійкості рівноваги систем багатьох частинок.

\keywords алгебра струмів, породжуючий функціонал Боголюбова, представлення колективних змінних, реконструкція оператора Гамільтона

\end{abstract}


\begin{thebibliography}{99}

\bibitem{Ar} Araki H., Publ. RIMS, Kyoto University, 1969/70, \textbf{5}, 361--422.


\bibitem{Aref} Arefyeva I.Ya., Teor. Mat. Fiz., 1972,
\textbf{10}, No.~2, 223--237 (in Russian).

\bibitem{CH} Carey A.L., Hannabuss K.C.,
%Temperature states on loop groups, theta functions and the Luttinger model.
J. Funct. Anal., 1987, \textbf{75}, 128--160; \doi{10.1016/0022-1236(87)90109-1}.

\bibitem{Newm} Newman Ch.M., Commun. Math. Phys., 1972, \textbf{26}, No.~3, 169--204; \doi{10.1007/BF01645089}.


\bibitem{Bom} Bohm D., The General Collective Variables Theory, Moscow, Mir,
1964 (in Russian).

%[5]
\bibitem{Bo} Bogoliubov N.N., Problems of Dynamic Theory in Statistical Physics,
OGIZ, Gostekhizdat, Moscow, 1946 (in Russian) [Bogoliubov N.N., Problems of Dynamic Theory in Statistical Physics, Technical Information Service, Oak Ridge, Tennessee,  1960].

\bibitem{Zub} Zubarev D.N., DAN SSSR, 1954, \textbf{95}, No.~4, 757--760 (in Russian).

\bibitem{Yu} Yukhnovskii I.R., Holovko M.F.,  Statistical Physics of
Equilibrium Systems, Kyiv, Naukova Dumnka, 1980 (in Russian).

\bibitem{BP} Bogolubov N.N. (Jr.), Prykarpatsky A.K.,
%The N.N. Bogolubov generating functional method in statistical mechanics and a collective
%variables transform analog.
Teor. Mat. Fiz., 1986, \textbf{66}, No.~3,
 463--480 (in Russian).

\bibitem{Pr} Prykarpatsky A.K.,
%The N.N. Bogolubov generating functional
%method in statistical mechanics and a collective variables transform analog
%within the grand canonical ensemble.
DAN AN SSSR, 1985, \textbf{285}, No.~5, 1096--1101 (in Russian).

\bibitem{BP1} Bogolubov N.N. (Jr.), Prykarpatsky A.K.,
%The Bogolubov quantum generating functional in statistical physics: Lie algebra of
%currents, its represenations and functional equations.
Phys. Part. Nuclei, 1986, \textbf{17}, No.~4, 789--920 (in Russian)

\bibitem{Vl} Vladimirov V.S., Generalized Functions in Mathematical Physics,
Moscow,  Nauka, 1979, (in Russian).

%[10]
\bibitem{MS} Goldin G.A., Menicoff R., Sharp D.H., J. Math. Phys., 1980, \textbf{21}, No.~4, 650--664; \doi{10.1063/1.524510}.

%[9]
\bibitem{MS1} Goldin G.A., Grodnik J., Powers R.T., Sharp D.H.,
J. Math. Phys., 1974, \textbf{15}, No.~1, 88--100; \doi{10.1063/1.1666513}.


%[14]
\bibitem{BB} Bogolubov N.N., Bogolubov N.N. (Jr.), Introduction into
Quantum Statistical Mechanics, World Scientific, New Jersey, 1986.

\bibitem{BS} Bogolubov N.N., Shirkov D.V., Introduction to the Theory of
Quantized Fields, Interscience, New York, 1959.

%%%

\bibitem{Leb1} Aizenman M., Goldstein S., Gruber C., Lebowitz J.L.,
Martyin P.,
%On the equivalence between KMS-states and equilibrium states for classical systems.
Commun. Math. Phys., 1977, \textbf{53}, 209--220; \doi{10.1007/BF01609847}.

%\bibitem{BSh} Bogolubov N.N., Shirkov D.V., Introduction to the Theory of
%Quantizerd Fields, Interscience, New York, 1959.

\bibitem{Leb} Lebowitz J.L., Aizenman M., Goldstein S.,
%On the stability of equilibrium states of finite classical systems.
J. Math. Phys., 1975, \textbf{16}, No.~6, 1284--1287; \doi{10.1063/1.522681}.



\end{thebibliography}
\end{document}